\documentclass[3p,10pt,a4paper,twoside,fleqn,sort&compress]{elsarticle}
\makeatletter
\def\ps@pprintTitle{%
  \let\@oddhead\@empty
  \let\@evenhead\@empty
  \let\@oddfoot\@empty
  \let\@evenfoot\@oddfoot
} \makeatother

\usepackage{graphicx}
\usepackage{verbatim}
\usepackage{amsmath}
\usepackage[varg]{pxfonts}
\usepackage{eepic}
\usepackage{amssymb}

\usepackage{mathtools}
\usepackage[all,cmtip]{xy}

\usepackage{tikz}
\usetikzlibrary{matrix,arrows}
\usepackage{pdflscape, xtab,longtable}

\newtheorem{theorem}{Theorem}[section]

\newtheorem{remark}[theorem]{Remark}

\begin{document}

\begin{frontmatter}

\title{\Large\bf A unified characterization of \\ generalized information and certainty measures\tnoteref{t1}}
\tnotetext[t1]{Research supported by Ministry  of Science and
Technological Development, Republic of Serbia, Grants No. 174013
and 174026}

\author[misanu]{Velimir M. Ili\'c\corref{cor}}
\ead{velimir.ilic@gmail.com}

\author[fosun]{Miomir S. Stankovi\'c}
\ead{miomir.stankovic@gmail.com}

\cortext[cor]{Corresponding author. Tel.: +38118224492; fax:
+38118533014.}
\address[misanu]{Mathematical Institute of the Serbian Academy of Sciences and Arts, Kneza Mihaila 36, 11000 Beograd, Serbia}
\address[fosun]{University of Ni\v s, Faculty of Occupational Safety, \v Carnojevi\'ca 10a, 18000 Ni\v s, Serbia}

\begin{abstract}
In this paper we consider the axiomatic characterization of
information and certainty measures in a unified way. We present
the general axiomatic system which captures the common properties
of a large number of the measures previously considered by
numerous authors. We provide the corresponding characterization
theorems and define a new generalized measure called the Inforcer,
which is the quasi-linear mean of the function associated to the
event probability following the general composition law.
In particular, we pay attention to the polynomial composition and
the corresponding polynomially composable Inforcer measure. The
most common measures appearing in literature can be obtained by
specific choice of parameters appearing in our generic measures
and they are listed in tables.

\end{abstract}
\begin{keyword}
information measure; entropy;  inaccuracy; certainty; axiomatic
characterization; pseudo-addition; polynomial addition
\end{keyword}

\end{frontmatter}

\newcommand{\bs}{\boldsymbol}
\newcommand{\mc}{\mathcal}
\newcommand{\comp}{\circ}

\newcommand{\mcP}{P}
\newcommand{\mcQ}{Q}
\newcommand{\mcU}{U}
\newcommand{\mcV}{V}
\newcommand{\mcPQ}{\mcP\mcQ}

\newcommand{\mcA}[1]{{\mc G}_{#1}}
\newcommand{\mcAn}[1]{\bar{\mc G}_{#1}}
\newcommand{\iec}{\mc{G}}

\newcommand{\mcH}{{\mc H}}

\newcommand{\mcHn}[2]{{\mbox{H}}_{#1}^{#2}}
\newcommand{\Cpl}[2]{\mbox{C}_{#1}^{#2}}

\newcommand{\mcI}[2]{\mc I_{#1}^{#2}}
\newcommand{\mcC}[2]{\mc C_{#1}^{#2}}

\newcommand{\ec}{{\mc E\mc C}}
\newcommand{\cc}{{\mc C\mc C}}
\newcommand{\ic}{\mc{EI}}
\newcommand{\e}{2}

\newcommand{\Logp}{\hbox{Log}_\op}
\newcommand{\Logq}{\hbox{Log}_q}
\newcommand{\Expp}{\hbox{Exp}_\op}
\newcommand{\op}{\oplus}
\newcommand{\od}{\odot}
\newcommand{\ot}{\otimes}
\newcommand{\cop}{\tau}
\newcommand{\isoXtoR}{h^{-1}}
\newcommand{\isoRtoXcer}{h_C}
\newcommand{\isoXtoRcer}{h_C^{-1}}
\newcommand{\isoRtoXinf}{h_I}
\newcommand{\isoXtoRinf}{h_I^{-1}}
\newcommand{\isoRtoX}{h}
\newcommand{\Jp}{\mc J}

\newcommand{\ra}{\rightarrow}

\newcommand{\sR}{\mathbb R}
\newcommand{\sN}{\mathbb N}
\newcommand{\sX}{\mathbb X}
\newcommand{\pvec}{(p_1, \dots, p_n)}
\newcommand{\qvec}{(q_1, \dots, q_m)}
\newcommand{\escP}{\mcP^{(\alpha)}}
\newcommand{\escPau}{\mcP^{(\alpha)}}
\newcommand{\escpvec}{( \escp{1}, \dots, \escp{n})}
\newcommand{\escp}[1]{\frac{p_{#1}^{(\alpha)}}{\sum_{i=1}^n
p_i^{(\alpha)}}}
\newcommand{\escpau}[1]{u_{#1}p_{#1}^{\alpha}}
\newcommand{\escr}[1]{r_{#1}^{(\alpha)}}
\newcommand{\escq}[1]{q_{#1}^{(\alpha)}}
\newcommand{\escqc}[2]{q_{#1 | #2}^{(\alpha)}}

\newcommand{\sgn}{\mbox{sgn}}

\newcommand{\uw}[1]{u_#1}
\newcommand{\vw}[1]{v_#1}

\newcommand{\rabi}{\ra}

\newcommand{\twoon}[1]{\exp_2\left\{#1\right\}}

\newcommand{\Itil}{\widetilde{\mc I}}
\newcommand{\Ctil}{\widetilde{\mc C}}
\newcommand{\gtil}{\widetilde{g}}

\section{Introduction}

In the past decades there was a plausible interest for definition
and characterization of the measures of certainty and information
associated to a probability distribution. The most commonly used
ones are those which can be obtained as the average value of the
informaion/certainty associated to the event.

Information measures determine the amount of uncertainty
associated to a probability distribution. The basic one is the
Shannon entropy \cite{shannon_48}, defined as a linear
(trace-form) expectation of an additive decreasing function of an
event probability called information content. Ren\'yi
\cite{Renyi_70}, Varma \cite{Varma_66} and Nath \cite{Nath_68}
considered the class of entropies which can be obtained as
quasi-linear mean weighted by the random variable probability.
Information entropies with more general weights were considered by
Acz\'el and Dar\'oczy \cite{Aczel_Daroczy_63}, Kapur
\cite{Kapur_83}, Rathie \cite{Rathie_70}, Khan and Autar
\cite{Khan_Autar_79} and Singh et. al \cite{Singh_et_al_03}.
Havrda and Charv\'at \cite{Havrda_Charvat_67} and, subsequently,
Dar\'oczy \cite{Daroczy_70} and Tsallis \cite{tsallis_88},
considered the entropies which are the trace form of
{pseudo-additive} information content. The trace form entropies
based on the pseudo-additive content are also considered by Abe
\cite{Abe_97} and Kaniadakis \cite{Kaniadakis_02a},
\cite{Kaniadakis_05}.
The class of entropies which are quasi-linear mean of the
pseudo-additive information are considered by Sharma and Mittal
\cite{Sharma_Mittal_75}, Frank and
Daffertshofer\cite{Frank_Daffertshofer_00}, Arimoto
\cite{Arimoto_71}, Boekee, Boxma and Van der Lubbe
\cite{Boekee_Lubbe_80}, \cite{Lubbe_et_al_84} and Picard
\cite{picard_79}.

Inaccuracy measures are a generalization of information entropies,
which deal with two distributions and reduce to the entropies if
the distributions are identical. The firstly introduced one was
the Kerridge inaccuracy \cite{Kerridge_61}, defined as the
expected value of the information content of the first
distribution, where the weights are are event probabilities with
respect to the second distribution. Nath \cite{Nath_70} considered
two types of the generalization - one to quasi-linear means and
the other with pseudo-additive information content. The
combination of these two approaches was considered by Gupta and
Sharma \cite{Gupta_Sharma_76}.

Certainty measures are defined as the average value of a
multiplicative increasing function of the event probability called
certainty content. The certainty measures which can be represented
as the trace form expectation of the certainty content are
Onicesu's information energy \cite{Onicescu_66}, also called
Weaver's expected commonness \cite{Weaver_66}, and order-$\alpha$
weighted information energy introduced by Pardo
\cite{Pardo_86}.~The certainty measures which can be represented
as the quasi-linear expectation were considered by Van der Lubbe
\cite{Lubbe_et_al_84} and Bhatia \cite{Bhatia_10}.

In this paper we propose the general axiomatic system, which
characterizes in a unique manner the majority of the known
information and certainty measures, and obtain the Inforcer
measure
as the unique one that satisfies it. By our axiomatic system,
information and certainty measures are the particular cases of the
Inforcer measure, which is the quasi-linear of the Inforcer
content. The Inforcer content is defined as a monotonic function
of event probability, having the information and certainty content
as special cases. The Inforcer measure and the Inforcer content
follow the simple composition rule which asserts that the Inforcer
measure and the Inforcer content of joint distributions can be
obtained as the composition of the Inforcer measure and the
Inforcer content of particular distributions.

In particular, we pay attention to the most common measures
appearing in literature, generated by the polynomial composition
operation. The polynomial composition has already been considered
by Behara and Nath \cite{Behara_Nath_74} and by Ebanks
\cite{Ebanks_84} for the case of trace form information entropies.
Instead of this, we derive the more general form for information
and certainty measures generated as a quasi-linear mean of the
content which follows the polynomial composition law.
%
It is shown that pseudo-addition \cite{Nivanen_Wang__03} and real
product represent the unique polynomial composition operations
which preserve the decreasing information and increasing certainty
content.
Finally, we give the theorem which represents the interplay
between the polynomial certainty and information measures
generalizing the result from \cite{Lubbe_et_al_84}.

The general axiomatic system for the Inforcer measure and the
corresponding uniqueness theorem are presented in section
\ref{sec: axioms}. In section \ref{sec: inf and cert} we define
the information and certatinty measures as instances of Inforcer
measure and derive relation between them. In section \ref{sec:
poly content} we consider polynomial composition and derive the
corresponding information and certainty measures.~The majority of
the information and certainty measures previously considered in
literature are obtained by instantiation of the generalized
average content and listed in Tables \ref{tableEntropy} and
\ref{tableCertainty}.

\section{Axiomatic characterization of the Inforcer measure}
\label{sec: axioms}

Let $\isoRtoX: \sR \ra \sR$ be a monotonic continuous (hence
invertible) function such that $h(x)>0$ for $x>0$ and let the
composition operation $\od$ be defined as:
\begin{equation}
\isoRtoX(a + b) = \isoRtoX(a) \od \isoRtoX(b); \quad a, b \in \sR.
\end{equation}
Let the set of all $n$-dimensional distributions and the set of
all the positive ones be denoted with
\begin{equation}
   \Delta_n \equiv \left\{ (p_1, \dots , p_n) \Big\vert \; p_i \ge 0,
      \sum_{i=1}^{n} p_i = 1 \right \};\quad
   \Delta_n^+ \equiv \left\{ (p_1, \dots , p_n) \Big\vert \; p_i > 0,
      \sum_{i=1}^{n} p_i = 1 \right \};\quad  n>1,
  \label{Delta}
\end{equation}
respectively. Let the \textit{direct product}, $\mcP \star \mcQ
\in \Delta_{nm}$, be defined as
\begin{equation}
\mcP \star \mcQ = (p_1q_1, p_1q_2, \dots, p_n q_m),
\end{equation}
for $\mcP = \pvec \in \Delta_n$ and $\mcQ = \qvec \in \Delta_m$
let and $\sR^+$ denotes the set of positive real numbers.

The Inforcer measure is characterized with the following set of
axioms.

\begin{description}
\item[{[A1]}]
The \emph{Inforcer content} $\ec: (0,1] \ra \sR^+$ is a continuous
monotonic function, which is composable:
\begin{equation}
\label{Reny 1: axioms: I additivity} \ec(p q) = \ec(p) \od \ec(q),
\quad \text{for all} \quad p,q \in (0,1].
\end{equation}

\item[{[A2]}] The $n$-dimensional Inforcer measure is a continuous function, $\iec: \Delta_n \times \Delta_n \ra
\sR^+$, $n\in\sN$, which to every pair of distributions
 $\mcU=(u_1,\dots,u_n) \in \Delta_n$ and $\mcP=(p_1,\dots,p_n)\in
\Delta_n^+$ assigns a quasi-linear mean of the content
\begin{equation}
\label{Reny 1: axioms: A is mean}%
\iec(\mcU; \mcP) = f^{-1}\left( \sum_{k=1}^{n} \uw{k} \cdot
f(\ec(p_k) \right),
\end{equation}
where $f$ is invertible and continuous.%

\item[{[A3]}] $\iec$ is composable:
\begin{equation}
\label{Reny 1: axioms: A additivity}%
\iec(\mcU \star \mcV ; \mcP \star \mcQ) = \iec(\mcU; \mcP) \od
\iec(\mcV; \mcQ),
\end{equation}
for all $\mcU,\mcV,\mcP, \mcQ \in \Delta_n$.
\end{description}

The following theorem gives the unique class of functions to which
the Inforcer measure belongs.

\begin{theorem}\rm
\label{Reny 1: theorem}

Let the axioms [\textbf{A1}]-[\textbf{A3}] hold. Then, the Inforcer content and the Inforcer measure are uniquely determined with 
\begin{equation}
\label{uniq theo: ec} \ec(p)= \isoRtoX(\cop \cdot \log_2 p), \quad
\cop < 0
\end{equation}
and
\begin{equation}
\label{sec: iec: iec form}
\iec(\mcU; \mcP) =%
\begin{dcases}
 \mcA{\cop}(\mcU; \mcP) = \isoRtoX \left( \sum_{k=1}^{n} \uw{k} \ \log_2
p_k^\cop \right) , \quad &\lambda = 0 \\
 \mcA{\cop,\lambda}(\mcU; \mcP) = \isoRtoX \left( \frac{1}{\lambda} \log_2 \left(\sum_{k=1}^{n} \uw{k} \
p_k^{\cop \lambda}\right) \right), \quad &\lambda \neq 0
\end{dcases}
\quad \text{where} \quad \cop<0.
\end{equation}

\end{theorem}

\textbf{Proof.}
Let $\comp$ denote the composition of functions and let $\mc J =
\isoXtoR \comp \ec$. By applying isomorphism $\isoXtoR$ to
(\ref{Reny 1: axioms: I additivity}), we get the Cauchy functional
equation,
\begin{equation}
\label{Reny 1: theorem: J aditivity} \Jp(pq) = \Jp(p) + \Jp(p),
\end{equation}
which has the unique solution
\begin{equation}
\Jp (p) = \cop \cdot \log_2 p \quad \Leftrightarrow \quad
\ec(p)=\isoRtoX(\Jp(p))=\isoRtoX(\cop \cdot \log_2 p),
\end{equation}
which proves the equation (\ref{uniq theo: ec}), since
the positivity of $\isoRtoX$ 
implies that $\cop <0$.

If we set $g = f \comp \isoRtoX$ or, equivalently, $f = g \comp
\isoXtoR$, (\ref{Reny 1: axioms: A is mean}) can be transformed to
\begin{equation}
\label{uniq theo: proof: A class form} \iec (\mcU; \mcP)= \isoRtoX
\left( g^{-1}\left(\sum_{k=1}^{n} \uw{k} \ g\left(\Jp(p_k)
\right)\right) \right)= \isoRtoX \left( g^{-1}\left(\sum_{k=1}^{n}
\uw{k} \ g(\tau \cdot \log_2 p_k) \right) \right).
\end{equation}

The function $g$ can be determined using the pseudo-additivity of
entropy (\ref{Reny 1: axioms: A additivity}) which has the form

\begin{equation}
\isoRtoX \left( g^{-1}\left(\sum_{k=1}^{n}\sum_{l=1}^{m}
\uw{k}\vw{l} \ g\left(\Jp(p_k q_l)
\right) \right) \right)=
\isoRtoX \left( g^{-1}\left(\sum_{k=1}^{n} \uw{k} \
g\left(\Jp(p_k) \right) \right) \right) \od \isoRtoX \left(
g^{-1}\left(\sum_{l=1}^{m} \vw{l} \ g\left(\Jp(q_l) \right)
\right) \right).
\end{equation}

Let $\mcQ = (1/m, \dots, 1/m)$ be uniform and $\Jp(1/m) =\Jp$. 
By applying $\isoXtoR$ and after using $\Jp(pq) = \Jp(p) + \Jp(q)$, we get

\begin{equation}
g^{-1}\left(\sum_{k=1}^{n} \uw{k} \ g\left(\Jp(p_k) + \Jp
\right) \right) =
g^{-1}\left(\sum_{k=1}^{n} \uw{k} \ g\left(\Jp(p_k) \right)
\right) + \Jp
\end{equation}
or, if we set $g_{\mc J}(z) = g(z+\mc J)$,
\begin{equation}
g_\Jp^{-1}\left(\sum_{k=1}^{n} \uw{k} \
g_\Jp\left(\Jp(p_k)\right)\right) =
g^{-1}\left(\sum_{k=1}^{n} \uw{k} \ g\left(\Jp(p_k) \right)
\right).
\end{equation}
Since $g_{\mc J}$ and $g$ generate the same mean, $g_{\mc J}$ is
the linear function of $g$ \cite{Hardy_et_al_34} and we get the
equation
\begin{equation}
\label{Reny 1: theorem: g(x) equation} g_{\mc J}(z) = a(\mc J)
\cdot g(z) + b(\mc J).
\end{equation}
The functional equation (\ref{Reny 1: theorem: g(x) equation}) can
be solved as in \cite{ Hardy_et_al_34}. Since $g(z+\mc J)=g(\mc
J+z)$, we can write $a(\mc J) g(z) +g(\mc J)=a(z) g(\mc J) +g(z) $
or $(a(\mc J)-1)/g(\mc J)=(a(z)-1)/g(z)=\gamma$. Inserting this
into equation (\ref{Reny 1: theorem: g(x) equation}) leads to the
following functional equations
\begin{alignat}{2}
g(\mc J+z) &= g(\mc J) + g(z) &\mbox{ for } \gamma=0 \\
a(\mc J+z) &= a(\mc J)\cdot a(z)  &\mbox{ for } \gamma\not=0.
\end{alignat}
The first case leads to $g(z)=cz$ whereas the second one imposes
to the function $a(z)=\e^{\lambda z}$.  Then, we can write $g(z) =
(2^{\lambda z}-1)/\gamma$, where $\lambda\neq0$, in this case
since $g$ must be invertible. The theorem is proven by
substitution of the solution for $g$ in (\ref{uniq theo: proof: A
class form}). $\square$

If $\mcU\equiv\mcP$ and the normalization condition $\mc G(
\frac{1}{2}, \frac{1}{2}; \mcU ) = \isoRtoX(1)$ is additionally
satisfied, then $\tau=-1$, and the class of Inforcer measures
(\ref{sec: iec: iec form}) reduces to the class derived and
characterized in \cite{Amblard_Vignat_06}
\begin{equation}
\mc G(\mcU; \mcP) =%
\begin{dcases}
\ \isoRtoX \left( - \sum_{k=1}^{n} p_k \ \log_2
p_k \right), \quad &\lambda = 0 \\
\ \isoRtoX \left( \frac{1}{\lambda} \log_2 \sum_{k=1}^{n} \ p_k^{1
- \lambda} \right), \quad &\lambda \neq 0.
\end{dcases}
\label{Renyi 2: H L(g)}
\end{equation}
Note that, although not explicitly mentioned in
\cite{Amblard_Vignat_06}, the proof from \cite{Amblard_Vignat_06}
requires continuity of the event and the average content, the
continuity of $\isoXtoR$, as well as the normalization condition
(compare section 2 from \cite{Amblard_Vignat_06} with the proof of
theorem \ref{Reny 1: theorem}).

\begin{remark}\rm
The structure $(\sR, \od)$ is a commutative topological group
isomorphic to $(\sR,+)$ and $\isoRtoX$ is an isomorphism from
$(\sR,+)$ to $(\sR,\od)$, i.e. $(\sR, \od)$ is Abelian, $\sR^2 \ra
\sX: (a, b) \ra a \od b$, and $\sR \ra \sR: a \ra \ominus a$ are
continuous. Here, for simplicity, we consider topological groups
defined over $\sR$. However, the results from theorem \ref{Reny 1:
theorem} are valid for an arbitrary topological group.
\end{remark}

\section{Information and certainty measures}
\label{sec: inf and cert}

Using the Inforcer content and the Inforcer measure, entropy and
certainty measures can be defined as follows.
Let $u:\Delta_n \ra \Delta_n$ be a continuous function such that
$u(\mcP)=(u_1(\mcP),\dots,u_n(\mcP)) \in \Delta_n$.

If $h:\sR \rabi \sR$ is increasing and $h(0)=0$, the Inforcer
content is a decreasing function $\ec:(0,1] \rabi \sR^+$ and is
called the \emph{information content}. The corresponding Inforcer
measure, $\mcI{}{}:\Delta_n\times\Delta_n\ra\sR^+$, is called the
\emph{generalized inaccuracy measure}. The function $\mcH:
\Delta_n \ra \sR^+$,$\mcH(\mcP)=\iec(u(\mcP); \mcP)$ is called the
\emph{generalized entropy}. The inaccuracy and the entropy are
both referred to as the \emph{information measures}.

If $h:\sR \rabi \sR^+$ is decreasing and $h(+\infty)=0$, the
Inforcer content is an increasing function $\ec:(0,1] \rabi (0,1]$
and is called the \emph{certainty content}. The corresponding
Inforcer measure, $\mcC{}{}: \Delta_n \times \Delta_n \ra \sR^+$,
%
%
%
is called the \emph{generalized certainty measure}.

Previously, the inaccuracy measures have been considered in
\cite{Kerridge_61}, \cite{Nath_70}, \cite{Gupta_Sharma_76},
entropies in \cite{shannon_48}, \cite{Renyi_70}, \cite{Varma_66},
\cite{Nath_68}, \cite{Aczel_Daroczy_63}, \cite{Kapur_83},
\cite{Rathie_70}, \cite{Khan_Autar_79}, \cite{Singh_et_al_03},
\cite{Havrda_Charvat_67}, \cite{Daroczy_70}, \cite{tsallis_88},
\cite{Abe_97}, \cite{Kaniadakis_02a}, \cite{Kaniadakis_05},
\cite{Sharma_Mittal_75}, \cite{Frank_Daffertshofer_00},
\cite{Arimoto_71}, \cite{Boekee_Lubbe_80}, \cite{Lubbe_et_al_84},
\cite{picard_79}, and certainty measures in \cite{Onicescu_66},
\cite{Weaver_66} \cite{Pardo_86}, \cite{Lubbe_et_al_84},
\cite{Bhatia_10}.
Most of these measures follow the polynomial composition law,
which will be discussed in section \ref{sec: poly content}.

With following theorem we establish the connection between
information and certainty measures, by which an information
measure is uniquely determined as a decreasing function of a
certainty measure. The theorem generalizes the result from
\cite{Lubbe_et_al_84}, which relates the pseudo-additive entropies
and multiplicative certainty measures.

\begin{theorem}\rm
\label{rel cer inf: theorem}

Let $\mcP = (p_1, \dots, p_n) \in \Delta_n$, $\mcQ = (q_1, \dots,
q_m) \in \Delta_m$ $n, m \in \sN$ and let $\mcC{}{}:
\Delta_n\times \Delta_n \ra \sR^+$ be a certainty measure and a
function $\mcI{}{}: \Delta_n\times\Delta_n \ra \sR^+$ have the
following properties:

\begin{description}

\item[{[P1]}] $\mcI{}{}$ is the pseudo-additive measure
\begin{equation}
\label{rel cer inf: H(PQ) =H(P) * H(Q)} \mcI{}{}(\mcP \star \mcQ;
\mcU \star \mcV)=\mcI{}{}(\mcU; \mcP) \op \mcI{}{}(\mcQ; \mcV);
\end{equation}
where $\op$ is defined with $\isoRtoXinf(a + b) = \isoRtoXinf(a)
\op \isoRtoXinf(b)$ for all $a, b \in \sR$ and $\isoRtoXinf: \sR
\ra \sR^+$ is an increasing function, such that $h(0)=0$.

\item[{[P2]}] $\mcI{}{}$ is a continuous and strictly
monotonic function $g:\sR\ra\sR$ of the certainty measure
$\mcC{}{}$, so that
\begin{equation}
\label{rel cer inf: H =g(C)}
\mcI{}{}(\mc P; \mcU) = g(\mcC{}{}(\mc P; \mcU)).
\end{equation}

\end{description}
Thus, $\mcI{}{}$ is the generalized inaccuracy measure:
\begin{equation}
\mcI{}{}(\mcU; \mcP) =%
\begin{dcases}
\ \isoRtoXinf \left( \sum_{k=1}^{n} \uw{k} \ \log_2
p_k^\cop \right) , \quad &\lambda = 0 \\
\ \isoRtoXinf \left( \frac{1}{\lambda} \log_2 \left(\sum_{k=1}^{n}
\uw{k} \ p_k^{\cop \lambda}\right) \right), \quad &\lambda \neq 0
\end{dcases}
\quad \text{where} \quad \cop<0.
\end{equation}

\end{theorem}

\textbf{Proof:} By the definition, a certainty measure is an
Inforcer measure if the function $h\equiv\isoRtoXcer
: \sR \ra \sR^+$ is a decreasing function and
$\isoRtoXcer(+\infty)=0$, so that we have
\begin{equation}
\label{rel cer inf: C form}
\mcC{}{}(\mcU; \mcP) =%
\begin{dcases}
\ \isoRtoXcer \left( \sum_{k=1}^{n} \uw{k} \ \log_2
p_k^\cop \right) , \quad &\lambda = 0 \\
\ \isoRtoXcer \left( \frac{1}{\lambda} \log_2 \left(\sum_{k=1}^{n}
\uw{k} \ p_k^{\cop \lambda}\right) \right), \quad &\lambda \neq 0
\end{dcases}
\quad \text{where} \quad \cop<0.
\end{equation}

We will show that $g(y)=\isoRtoXinf(k\cdot \isoXtoRcer(y))$ and
the result follows from [\textbf{P2}]. Note that $k > 0 $ since
$\mcI{}{}$ is by assumption positive, $\isoRtoXinf$ is increasing,
and $\isoRtoXcer$ is decreasing, so we can fix the value of the
$k$ to $1$ since the $\cop$ and $\lambda$ in (\ref{rel cer inf: C
form}) can be arbitrarily chosen.

Let us denote $\Itil=\isoXtoRinf \comp \mcI{}{}$, $\
\Ctil=\isoXtoRcer \comp \mcC{}{}$, and $\gtil = \isoXtoRinf \comp
g \comp \isoRtoXcer$. Since $\mcI{}{}=g \comp \mcC{}{}$, we have
$\Itil=\gtil\comp\Ctil$, or
\begin{equation}
\label{rel cer inf: gtil(ctil)=itil}
\gtil\left(\Ctil(\mcP \star \mcQ; \mcU \star \mcV)\right)=%
\Itil(\mcP \star \mcQ; \mcU \star \mcV)=%
\Itil(\mcU; \mcP) + \Itil{}{}(\mcQ; \mcV),
\end{equation}
where the right-hand side equality follows from (\ref{rel cer inf:
H(PQ) =H(P) * H(Q)}).
As the special case of the Inforcer measure, $\mcC{}{}$ satisfies
the composability axiom [\textbf{A3}]
\begin{equation}
\label{rel cer inf: C(PQ) =C(P) * C(Q)} \mcC{}{}(\mcP \star \mcQ;
\mcU \star \mcV)=\mcC{}{}(\mcU; \mcP) \ot \mcC{}{}(\mcQ; \mcV),
\end{equation}
where $\ot$ is defined with $\isoRtoXcer(a + b) = \isoRtoXcer(a)
\ot \isoRtoXcer(b)$ for all $a, b \in \sR$, and we have
\begin{equation}
\label{rel cer inf: ctil= ctil + ctil} \Ctil(\mcP \star \mcQ; \mcU
\star \mcV)=\Ctil(\mcU; \mcP) + \Ctil{}{}(\mcQ; \mcV).
\end{equation}
By combining the equations (\ref{rel cer inf: gtil(ctil)=itil})
and (\ref{rel cer inf: ctil= ctil + ctil}) we get
\begin{equation}
\gtil\left( \Ctil(\mcU; \mcP) + \Ctil{}{}(\mcQ; \mcV) \right)=
\Itil(\mcU; \mcP) + \Itil{}{}(\mcQ; \mcV).
\end{equation} By using
$\Itil=\gtil\comp\Ctil$ and setting $a=\Ctil(\mcU; \mcP)$,
$b=\Ctil{}{}(\mcQ; \mcV)$, we get the Cauchy functional equation
$\gtil(a)+\gtil(b)=\gtil(a+b)$, which has the unique solution
$\gtil(x)=k \cdot x$, $k \in \sR$ \cite{aczel1966lectures}.
Accordingly, $\gtil(x) = \isoXtoRinf\left( g \left(
\isoRtoXcer(x)\right)\right) = k x$ or equivalently, $ g \left(
\isoRtoXcer(x)\right) = \isoRtoXinf(kx)$. Finally, if we set $y =
\isoRtoXcer(x)$, we get $g(y)=\isoRtoXinf(k\cdot \isoXtoRcer(y))$
and the theorem is proven. $\square$

\section{The polynomial composable Inforcer measures}
\label{sec: poly content}

In this section we consider the Inforcer measures which follow the
polynomial composition law. More specifically, we consider the
case when the operation $\od$ can be represented as
\begin{equation}
\label{sec: h choice: add theo def} h(x+y) = h(x) \op h(y) =
F(h(x), h(y))
\end{equation}
and the function $F:\sR^2 \ra \sR$ is a two-variable polynomial.

The equation of the type (\ref{sec: h choice: add theo def}) is
called a polynomial addition theorem. In \cite{aczel1966lectures},
it is shown that the most general functions with a polynomial
addition theorem are
\begin{equation}
\label{sec: h choice: linear add theo sol} h(x) = a \cdot x + b,
\quad
\text{and} \quad h(x) = \frac{2^{c x} - d}{e},
\end{equation}
where $e \neq 0$, $a,b,c,d$ are arbitrary constants.
The respective polynomials $F$ in (\ref{sec: h choice: add theo
def}) are
\begin{equation}
F(u, v) = u + v + b \quad \text{and} \quad F(u, v) = euv + du + dv
+ \frac{d^2 - d}{e}.
\end{equation}
Note that the first formula reduces to real addition for $b=0$ and
the second one reduces to multiplication for $e=1$ and $d=0$.

The formulas (\ref{sec: iec: iec form}) and (\ref{sec: h choice:
linear add theo sol}) determine the most general form of
polynomially composable Inforcer measure. The values of the
parameters are further restricted if the Inforcer is considered as
an information or a certainty measure.
Previously, the polynomial composition has been considered by
Behara and Nath \cite{Behara_Nath_74} and by Ebanks
\cite{Ebanks_84}, for the case of trace form information measures.
In the following subsection, we derive the more general form for
information measures generated as a quasi-linear mean of the
content. Subsequently, we consider the polynomially composable
certainty measures.

\subsection{Polynomially composable information measures}

In the case of information measures, by definition, $h:\sR \rabi
\sR$ is an increasing function such that $h(0)=0$, which implies
$b=0$, $a>0$ and $d=1$, $c\cdot e>0$ in (\ref{sec: h choice:
linear add theo sol}), so we have 
\begin{equation}
\label{sec: h choice: inf: h}
\isoRtoX(x) = %
\begin{dcases}
\quad ax,\quad\quad\quad a>0,
\quad &\mbox{ for } e = 0 \\
\frac{2^{c \cdot x} - 1}{e},\quad c\cdot e > 0, \quad &\mbox{ for
} e \neq 0.
\end{dcases}
\end{equation}
The corresponding composition operation is defined with
\begin{equation}
\label{sec: h choice: inf: oplus} F(u, v) = u \op_e v = u + v
\quad \text{and} \quad F(u, v) =u \op_e v = u + v +
e
uv,
\end{equation}
which is the pseudo-addition defined in
\cite{Nivanen_Wang__03}, \cite{Borges_04}.

Using the pseudo-addition (\ref{sec: h choice: inf: oplus}) and
the form of function $h$ given by (\ref{sec: h choice: inf: h}),
polynomially additive information measures can be characterized
with the following instance of the axiomatic system
[\textbf{A1}]-[\textbf{A3}].

\begin{description}
\item[{[I1]}]
The \emph{information content} $\ic: (0,1] \ra \sR^+$ is a
continuous decreasing function, which is pseudo-additive:
\begin{equation}
\label{Reny 1: axioms: I additivity} \ic(p q) = \ic(p) \op_e
\ic(q), \quad \text{for all} \quad p,q \in (0,1].
\end{equation}

\item[{[I2]}] The $n$-dimensional {inaccuracy measure} is a continuous function, $\mcI{}{}: \Delta_n \times \Delta_n \ra
\sR^+$, $n\in\sN$, which to every pair of distributions
 $\mcU=(u_1,\dots,u_n) \in \Delta_n$, $\mcP=(p_1,\dots,p_n)\in
\Delta_n^+$ assigns a quasi-linear mean of the information content
\begin{equation}
\mcI{}{}(\mcU; \mcP) = f^{-1}\left( \sum_{k=1}^{n} \uw{k} \cdot
f(\ic(p_k) \right),
\end{equation}
where $f$ is invertible and continuous.%

\item[{[I3]}] $\mcI{}{}$ is pseudo-additive:
\begin{equation}
\mcI{}{}(\mcU \star \mcV ; \mcP \star \mcQ) = \mcI{}{}(\mcU; \mcP)
\od \mcI{}{}(\mcV; \mcQ),
\end{equation}
for all $\mcU,\mcV,\mcP, \mcQ \in \Delta_n$.
\end{description}

According to formulas (\ref{sec: iec: iec form}) and (\ref{sec: h
choice: inf: h}), the generalized inaccuracy measures satisfying
[\textbf{I1}]-[\textbf{I3}] are uniquely determined with the class
\begin{equation}
\label{sec: pol cont: entropy: gen form}
\mcI{}{}(\mcU; \mcP) =%
\begin{dcases}
 \mcI{\cop}{}(\mcU; \mcP) = \sum_{k=1}^{n} \uw{k} \ \log_2
p_k^\cop , \quad &\lambda = 0,\quad e=0 \\
 \mcI{\cop, \lambda}{}(\mcU; \mcP) = \frac{1}{\lambda} \log_2 \left( \sum_{k=1}^{n} \uw{k} \
p_k^{\cop \lambda}\right), \quad &\lambda \neq 0,\quad e=0 \\
%
\mcI{\cop}{c, e}(\mcU; \mcP) = \frac{1}{e} \left( \twoon{\sum_{k=1}^n {\cop \cdot c \cdot \uw{k}} \log p_k} - 1 \right) , \quad &\lambda = 0,\quad e \neq 0\\
 \mcI{\cop,\lambda}{c, e}(\mcU; \mcP) = \frac{1}{e} \left( \left(\sum_{k=1}^{n} \uw{k} \
p_k^{\cop \lambda}\right)^{\frac{c}{\lambda}} - 1\right), \quad
&\lambda \neq 0,\quad  e  \neq 0
\end{dcases}
\end{equation}
where $\cop<0$ and $c \cdot e>0$. Note that, in the case of $e=0$,
we can fix the value of the parameter $a$ appearing in the
function (\ref{sec: h choice: inf: h}) to $a=1$, since $a$ is a
multiplicative term and $\cop$ and $\lambda$ in (\ref{sec: pol
cont: entropy: gen form}) can be arbitrarily chosen.

If the normalization condition $\mcI{}{}( \frac{1}{2},
\frac{1}{2}; \mcU ) = 1$ is additionally satisfied, then
$\tau=-1$, and $e=2^c-1$ and the class of inaccuracy measures
(\ref{sec: pol cont: entropy: gen form}) reduce to the class
derived and characterized in \cite{Gupta_Sharma_76}. The majority
of previously considered information measures obtainable by a
specific choice of the parameters in (\ref{sec: pol cont: entropy:
gen form}) are listed in the Table \ref{tableEntropy}.

\subsection{Polynomially composable certainty measures}

In the case of certainty measures, by definition, $h:\sR \rabi
\sR^+$ is a positive decreasing function such that $h(+\infty)=0$.
The linear function case is not possible since a linear decreasing
function cannot be positive for all $x \in \sR^+(0,\infty)$. In
the case of the exponential function, $d=0$ since $h(+\infty)=0$.
Accordingly, for certainty measures, (\ref{sec: h choice: linear
add theo sol}) has the following form
\begin{equation}
\label{sec: h choice: cer: h}
\isoRtoX(x) = %
\frac{2^{-c \cdot x}}{e}; \quad c\cdot e> 0.
\end{equation}
The corresponding composition operation is defined with
\begin{equation}
\label{sec: h choice: cer: oplus} x \od y =%
e\cdot x \cdot y.
\end{equation}

Using the composition operation (\ref{sec: h choice: cer: oplus})
and the corresponding isomorphism (\ref{sec: h choice: cer: h})
the polynomial certainty measure can be characterized with the
following instance of the axiomatic system
[\textbf{A1}]-[\textbf{A3}].

\begin{description}
\item[{[C1]}]
The \emph{certainty content} $\cc: (0,1] \ra (0,1]+$ is a
continuous increasing function, which is multiplicative:
\begin{equation}
\label{Reny 1: axioms: I additivity} \cc(p q) = e \cdot \cc(p)
\cdot \cc(q), \quad \text{for all} \quad p,q \in (0,1].
\end{equation}

\item[{[C2]}] The $n$-dimensional {certainty measure} is a continuous function, $\mcC{}{}: \Delta_n \times \Delta_n \ra
\sR^+$, $n\in\sN$, which to every pair of distributions
 $\mcU=(u_1,\dots,u_n) \in \Delta_n$, $\mcP=(p_1,\dots,p_n)\in
\Delta_n^+$ assigns a quasi-linear mean of the certainty content
\begin{equation}
\mcC{}{}(\mcU; \mcP) = f^{-1}\left( \sum_{k=1}^{n} \uw{k} \cdot
f(\cc(p_k) \right),
\end{equation}
where $f$ is invertible and continuous.%

\item[{[C3]}] $\mcC{}{}$ is multiplicative:
\begin{equation}
\mcC{}{}(\mcU \star \mcV ; \mcP \star \mcQ) = \mcC{}{}(\mcU; \mcP)
\od \mcC{}{}(\mcV; \mcQ),
\end{equation}
for all $\mcU,\mcV,\mcP, \mcQ \in \Delta_n$.
\end{description}

According to formulas (\ref{sec: iec: iec form}) and (\ref{sec: h
choice: inf: h}) the generalized certainty measures satisfying
[\textbf{C1}]-[\textbf{C3}] are uniquely determined with the class
\begin{equation}
\label{sec: pol cont: certainty: gen form}
\mcC{}{}(\mcU; \mcP) =%
\begin{dcases}
\mcC{\cop}{}(\mcU; \mcP) = \frac{1}{e}\cdot
\twoon{- \sum_{k=1}^n {\cop \cdot c \cdot \uw{k}} \log p_k}, \quad
%
\quad &\lambda = 0,\ \gamma \neq 0\\
\mcC{\cop,\lambda}{}(\mcU; \mcP) =  \frac{1}{e}  \left(
\sum_{k=1}^{n} \uw{k} \ p_k^{\cop
\lambda}\right)^{-\frac{c}{\lambda}}, \quad &\lambda \neq 0,\
\gamma \neq 0
\end{dcases}
\end{equation}
where $\cop<0$ and $c\cdot e>0$.

If $e=1$ and $\mcU\equiv\mcP$, the axiomatic system
[\textbf{C1}]-[\textbf{C3}] and the class of certainty measures
(\ref{sec: pol cont: certainty: gen form}) reduces to the one
considered in \cite{Lubbe_et_al_84}, while the case of $\mcU
=(\escp{1}, \dots, \escp{n})$ is considered in \cite{Bhatia_10}.
The majority of previously considered certainty measures which can
be obtained by a specific choice of the parameters in (\ref{sec:
pol cont: certainty: gen form}) are listed in Table
\ref{tableCertainty}.

\section{Conclusion and further work}

In this paper we considered the axiomatic characterization of
information and certainty measures and derived the Inforcer
measure, which generalizes all of them. The definition of the
Inforcer measure paves the way for unification of generalized
divergence measures \cite{Kannappan_Rathie_74}, \cite{Mittal_75},
\cite{Behara_Nath_80}, \cite{Taneja_Kumar_04}, \cite{Taneja_05},
which should be explored further.

According to the axiomatic system the Inforcer is a composable
measure which can be represented as quasi-linear mean-value of
composable Inforcer content. The composition operation was defined
using the monotonic function $h : \sR \ra \sR$, which is
increasing for the information and decreasing for the certainty
measures. As the simplest case, we defined the class of
composition operations under the assumption that the operation
should have polynomial representation by use of the polynomial
addition theorem \cite{aczel1966lectures}.
%
The discussion can be further generalized by assuming the rational
or algebraic functions instead of the polynomial, in which case
the rational and algebraic addition theorems
\cite{aczel1966lectures} should be used.

In addition, it seems important to further generalize the Inforcer
axiomatic system and to derive the generalization of the Inforcer
measure which covers Abe \cite{Abe_97}, Kaniadakis
\cite{Kaniadakis_02a}, \cite{Kaniadakis_05} and
Sharma-Mittal-Taneja \cite{Sharma-Taneja_74}, \cite{Mittal_75}
entropies, which were not covered by our framework.

\newpage
\begin{center}
\begin{longtable}{| p{6.3cm} | p{0.8cm} p{0.8cm} p{1cm} p{1.2cm} p{0.8cm} p{1.2cm}|}
\caption{Generalized information measures (\ref{sec: pol cont:
entropy: gen form}), with
$u_k=\frac{\phi_k(p_k)}{\sum_{k=1}^n\phi_k(p_k)}$
}
    \label{tableEntropy}\\
    \hline
    &&&&&&\\
    \centering Measure 
    & type &$\phi_k(p)$ & $c$ & $e$ & $\cop$  & $\lambda$\\
    &&&&&&\\
    \hline
    \endfirsthead
    \caption{(continuied)}\\
    \hline
    &&&&&&\\
    \centering Measure 
    & type & $\phi_k(p)$ & $c$ & $e$ & $\cop$ & $\lambda$ \\
    &&&&&&\\
    \hline
    &&&&&&\\
    \endhead
    &&&&&&\\
    \hline
    \endfoot
    &&&&&&\\
    \hline
    \endlastfoot
    &&&&&&\\
    \centering Shannon  \cite{shannon_48}
    &&&&&&\\    &&&&&&\\
    \centering $-\sum_{i=1}^n{p_i\log \, p_i}$ &
    $\mcI{\cop}{}$
    & $p$ & $-$ & $-$ & $-1$ & $-$ \\ 
    &&&&&&\\
    \centering R\'enyi \cite{Renyi_70}&
    &&&&&\\    &&&&&&\\
    \centering $\dfrac{1}{1-\alpha}\log \Big(\sum_{i=1}^n{p^\alpha_i}\Big)$&
    $\mcI{\cop,\lambda}{}$
    & $p$& $-$ & $-$ & $-1$ & $1-\alpha$\\
    &&&&&&\\
    \centering Varma \cite{Varma_66}, Nath \cite{Nath_68b}
     &&&&&&\\ &&&&&&\\
    \centering $\dfrac{1}{\mu-\alpha}\log \left(\sum_{i=1}^n{p_i^{\alpha-\mu+1}}\right)$
    &$\mcI{\cop, \lambda}{}$ & $p$ & $-$ & $-$ &$-1$ & $\mu-\alpha$\\
    &&&&&&\\
    \centering $\dfrac{\mu}{\mu-\alpha}\log \left(\sum_{i=1}^n{p_i^{\alpha/\mu}}\right)$
    &$\mcI{\cop}{}$ & $p$ & $-$ & $-$ &$-1$ & $1-\dfrac{\alpha}{\mu}$ \\
    &&&&&&\\
    \centering Nath \cite{Nath_68}
     &&&&&&\\ &&&&&&\\
    \centering $\dfrac{1}{1-\alpha}\log \left(\sum_{i=1}^n{p_i^{\mu\alpha-\mu+1}}\right)$
    &$\mcI{\cop, \lambda}{}$ & $p$ & $-$ & $-$ &$-\mu$ & $1-\alpha$\\
    &&&&&&\\
    \centering $\dfrac{1}{1-\alpha}\log \left(\sum_{i=1}^n{p_i^{\alpha^\mu}}\right)$
    &$\mcI{\cop, \lambda}{}$ & $p$ & $-$ & $-$ &$\dfrac{\alpha^\mu-1}{1-\alpha}$ & $1-\alpha$ \\
    &&&&&&\\
    \centering Acz\'el and Dar\'oczy \cite{Aczel_Daroczy_63}&
    &&&&&\\    &&&&&&\\
    \centering $-\dfrac{\sum_{i=1}^n{p^\beta_i\log \, p_i}}{\sum_{i=1}^n{p^\beta_i}}$&
    $\mcI{\cop, \lambda}{}$
    & $p^\beta$& $-$ & $-$ &$-1$ & $-$\\
    &&&&&&\\
    \centering $\dfrac{1}{\beta - \alpha}\log
    \left(\dfrac{\sum_{i=1}^n{p^\alpha_i}}{\sum_{i=1}^n{p_i^\beta}}\right)$&
    $\mcI{\cop,\lambda}{}$
    & $p^\beta$& $-$ & $-$ &$-1$ & $\beta-\alpha$ \\
    &&&&&&\\
    \centering Kapur \cite{Kapur_83}&
    &&&&&\\    &&&&&&\\
    \centering $\dfrac{1}{1-\alpha}\log\left(\dfrac{\sum_{i=1}^n{p^{\alpha+\beta-1}_i}}{\sum_{i=1}^n {p^\beta}} \right)$&
    $\mcI{\cop,\lambda}{}$
    & $p^\beta$ & $-$ & $-$ &$-1$ & $1-\alpha$\\
    &&&&&&\\
    \centering Rathie \cite{Rathie_70}&
    &&&&&\\    &&&&&&\\
    \centering $\dfrac{1}{1-\alpha}\log\left(\dfrac{\sum_{i=1}^n{p^{\alpha+\beta_i-1}_i}}{\sum_{i=1}^n{p^{\beta_i}_i}}\right)$&
    $\mcI{\cop,\lambda}{}$
    & $p^{\beta_i}$ & $-$ & $-$ &$-1$ & $1-\alpha$\\
    \centering Khan and Autar \cite{Khan_Autar_79}&
    &&&&&\\    &&&&&&\\
    \centering $\dfrac{1}{1-\alpha}\log\left(\dfrac{\sum_{i=1}^n {p^{\alpha+\beta-1}_i}v_i}{\sum_{i=1}^n {p^\beta v_i}} \right)$&
    $\mcI{\cop,\lambda}{}$
    & $p^\beta v_i $ & $-$ & $-$ & $-1$ & $1-\alpha$\\
    &&&&&&\\
    \centering Singh et al. \cite{Singh_et_al_03}&
    &&&&&\\    &&&&&&\\
    \centering $\dfrac{1}{1-\alpha}\log\left(\dfrac{\sum_{i=1}^n {p^{\alpha\beta}_i}v_i}{\sum_{i=1}^n {p_i^\beta v_i}} \right)$&
    $\mcI{\cop,\lambda}{}$
    & $p^\beta v_i$ & $-$ & $-$ &$-\beta$& $1-\alpha$ \\
    &&&&&&\\
   \centering Havrda and Charv\'at \cite{Havrda_Charvat_67}, Dar\'oczy \cite{Daroczy_70}&
    &&&&&\\    &&&&&&\\
    \centering $\dfrac{1}{2^{1-\gamma}-1}\left(\sum_{i=1}^n{p^{\gamma}_i}-1\right)$&
    $\mcI{\cop,\lambda}{c,e}$
    &     $p$& $1-\gamma$ & $2^{1-\gamma}-1$ &$-1$ & $1-\gamma$ \\
    &&&&&&\\
    \centering Sharma and Mittal \cite{Sharma_Mittal_75}&
    &&&&&\\    &&&&&&\\
    \centering $\dfrac{1}{2^{1-\gamma}-1}
    \left(\twoon{\sum_{k=1}^n (\gamma-1)\ p_k \log p_k} -
    1\right)$&
    $\mcI{\cop}{c,e}$
    &$p$& $1-\gamma$ & $2^{1-\gamma}-1$ & $-1$  & $-$\\
    &&&&&&\\
    \centering $\dfrac{1}{2^{1-\gamma}-1}\left[\left(\sum_{i=1}^n{p^{\alpha}_i}\right)^{1-\gamma\over 1-\alpha}-1\right]$&
    $\mcI{\cop,\lambda}{c,e}$
    &     $p$& $1-\gamma$ & $2^{1-\gamma}-1$ & $1-\alpha$ & $-1$\\
    &&&&&&\\
    \centering Tsallis \cite{tsallis_88}&
    &&&&&\\    &&&&&&\\
    \centering $\dfrac{1}{1-\gamma}\left(\sum_{i=1}^n{p^{\gamma}_i}-1\right)$&$\mcI{\cop,\lambda}{c,e}$
    &     $p$& $1-\gamma$ & $1-\gamma$ & $-1$ & $1-\gamma$ \\
    &&&&&&\\
    \centering Frank and Daffertshofer \cite{Frank_Daffertshofer_00}
    &&&&&&\\ &&&&&&\\
    \centering
    $\dfrac{1}{1-\gamma}\left(\twoon{\sum_{k=1}^n (\gamma-1)\
    p_k \log p_k} -
    1\right)$&
    $\mcI{\cop}{c,e}$
    &$p$& $1-\gamma$ & ${1-\gamma}$ & $-1$ & $-$\\
     &&&&&&\\
    \centering $\dfrac{1}{1-\gamma}\left[\left(\sum_{i=1}^n{p^{\alpha}_i}\right)^{1-\gamma\over 1-\alpha}-1\right]$&$\mcI{\cop,\lambda}{c,e}$
    &     $p$& $1-\gamma$ & ${1-\gamma}$ & $-1$ & $1-\alpha$ \\
    &&&&&&\\
    \centering Arimoto \cite{Arimoto_71}&
    &&&&&\\    &&&&&&\\
    \centering $\dfrac{1}{\gamma-1}\left(\left(\sum_{i=1}^n{p^{1/\gamma}_i}\right)^{\gamma}-1\right)$&$\mcI{\cop,\lambda}{c,e}$
    &     $p$& $\gamma-1$ & $\gamma-1$ & $-1$ & $\dfrac{\gamma-1}{\gamma}$ \\
    \newpage
    \centering Boekee and Van der Lubbe \cite{Boekee_Lubbe_80}&
    &&&&&\\    &&&&&&\\
    \centering $\dfrac{\gamma}{{1-\gamma}}\left(\left(\sum_{i=1}^n{p^{\gamma}_i}\right)^{1\over\gamma}-1\right)$&$\mcI{\cop,\lambda}{c,e}$
    &     $p$& $\dfrac{1-\gamma}{\gamma}$ & $\dfrac{1-\gamma}{\gamma}$ & $-1$ & $1-\gamma$ \\
    &&&&&&\\
    \centering Van der Lubbe et al. \cite{Lubbe_et_al_84}&
    &&&&&\\    &&&&&&\\
    \centering $\sum_{i=1}^n{p_i\log \, p_i^\cop}$&$\mcI{\cop}{}$
    & $p$ & $-$ & $-$ &$\cop$ & $-$\\ 
    &&&&&&\\
    \centering%
 $\dfrac{1}{\lambda} \log_2 \sum_{k=1}^{n} \ p_k^{\cop \lambda+1}$
%
    &$\mcI{\cop,\lambda}{}$
    &     $p$& $-$ & $-$ & $\cop$ & $\lambda$\\
    &&&&&&\\
    \centering
    $\dfrac{1}{e} \left(\twoon{\sum_{k=1}^n {\cop \cdot c \cdot
    p_k \log p_k}} - 1 \right)$
    %
    &$\mcI{\cop}{c,e}$
    &     $p$& $c$ & $e$ & $\cop$ & $-$\\
    &&&&&&\\
    \centering
    $\dfrac{1}{e} \left( \left(\sum_{k=1}^{n} \ p_k^{\cop
    \lambda+1}\right)^{\frac{c}{\lambda}} - 1\right)$
    &$\mcI{\cop,\lambda}{c,e}$
    &     $p$& $c$ & $e$ & $\tau$ & $\lambda$\\
    &&&&&&\\
    \centering Kerridge \cite{Kerridge_61}
    &&&&&&\\    &&&&&&\\
    \centering $-\sum_{i=1}^n{u_i\log \, p_i}$ &
    $\mcI{\cop}{}$
    & $u$ & $-$ & $-$ & $-1$ & $-$ \\ 
 &&&&&&\\
   \centering Nath \cite{Nath_70}&
    &&&&&\\    &&&&&&\\
    \centering $\dfrac{1}{2^{1-\gamma}-1}\left(\sum_{i=1}^n{u_i p^{\gamma-1}_i}-1\right)$&
    $\mcI{\cop,\lambda}{c,e}$
    &     $p$& $1-\gamma$ & $2^{1-\gamma}-1$ &$-1$ & $1-\gamma$ \\
    &&&&&&\\
    \centering $\dfrac{1}{1-\alpha}\log \Big(\sum_{i=1}^n u_i p^{\alpha-1}_i\Big)$&
    $\mcI{\cop,\lambda}{}$
    & $u$& $-$ & $-$ & $-1$ & $1-\alpha$\\
    &&&&&&\\
    \centering Gupta and Sharma \cite{Gupta_Sharma_76},  Picard \cite{picard_79}&
    &&&&&\\    &&&&&&\\
    \centering $\dfrac{1}{2^{1-\gamma}-1} \left( \twoon{\sum_{k=1}^n (\gamma - 1)\ u_k \log p_k} - 1 \right)$&
    $\mcI{\cop}{c,e}$
    &$u$& $1-\gamma$ & $2^{1-\gamma}-1$ & $-1$  & $-$\\
    &&&&&&\\
    \centering $\dfrac{1}{2^{1-\gamma}-1}\left[\left(\sum_{i=1}^n{u_i p^{\alpha-1}_i}\right)^{1-\gamma\over 1-\alpha}-1\right]$&$\mcI{\cop,\lambda}{c,e}$
    &     $u$& $1-\gamma$ & $2^{1-\gamma}-1$ & $1-\alpha$ & $-1$\\

    \end{longtable}

\end{center}

\newpage
\begin{center}
\begin{longtable}{| p{5.1cm} | p{0.8cm} p{0.8cm} p{1cm} p{1.2cm} p{1.2cm} p{1.2cm}|}
    \caption{Generalized certainty measures (\ref{sec: pol cont: certainty: gen form}), with $u_k=\frac{\phi_k(p_k)}{\sum_{k=1}^n\phi_k(p_k)}$}\label{tableCertainty}\\
    \hline
    &&&&&&\\
    \centering Measure 
    & type &$\phi_k(p)$ & $c$ & $e$ & $\cop$ & $\lambda$ \\
    &&&&&&\\
    \hline
    \endfirsthead
    \hline
    &&&&&&\\
    Measure 
    & type & $\phi_k(p)$ & $c$ & $e$ & $\cop$ & $\lambda$ \\
    &&&&&&\\
    \hline
    &&&&&&\\
    \endhead
    &&&&&&\\
    \hline
    \endfoot
    &&&&&&\\
    \hline
    \endlastfoot
    &&&&&&\\
    \centering Onicescu \cite{Onicescu_66}, Weaver \cite{Weaver_66}
    &&&&&&\\    &&&&&&\\
    \centering $\sum_{i=1}^n p^2_i$ &
    $\Cpl{\cop, \lambda}{}$
    & $p$ & $1$ & $1$  & $-1$ & $-1$ \\ 
    &&&&&&\\
    \centering Teodorescu \cite{Theodorescu_77}&
    &&&&&\\    &&&&&&\\
    \centering $\dfrac{1}{\gamma-1}\sum_{i=1}^n p^\gamma_i$
    &$\Cpl{\cop,\lambda}{}$
    & $p$& $\gamma-1$ & $\gamma-1$ & $-1$ & $1-\gamma$\\
    &&&&&&\\
    \centering  Pardo and Taneja \cite{Pardo_Taneja_93}&
    &&&&&\\    &&&&&&\\
     \centering $\sum_{i=1}^n p^\gamma_i$
    &$\Cpl{\cop,\lambda}{}$
    & $p$& $\gamma-1$ & $1$ & $-1$ & $1-\gamma$\\
    &&&&&&\\
    \centering  Pardo \cite{Pardo_86}&
    &&&&&\\    &&&&&&\\
    \centering $\dfrac{1}{\gamma-1}\dfrac{\sum_{i=1}^n u_i p^\gamma_i}{\sum_{i=1}^n u_i p_i}$
    &$\Cpl{\cop,\lambda}{}$
    & $u_k\cdot p$& $\gamma-1$ & $\gamma-1$ & $-1$ & $1-\gamma$\\
    &&&&&&\\
    \centering Tuteja et al. \cite{Tuteja_et_al_93}&
    &&&&&\\    &&&&&&\\
    \centering $\dfrac{1}{\gamma- 1} \left(\dfrac{\sum_{i=1}^n{u_i p^{\gamma}_i}}{\sum_{i=1}^n{u_i p_i}}\right)^{\gamma-1\over \beta-1}$
    &$\Cpl{\cop,\lambda}{c, e}$
    & $u_k\cdot p$& $\gamma-1$ & $\gamma-1$ & $\dfrac{\gamma-1}{1-\beta}$ & $1-\beta$\\
    &&&&&&\\
    \centering Van der Lubbe et al. \cite{Lubbe_et_al_84}&
    &&&&&\\    &&&&&&\\
    \centering $\sum_{k=1}^n \twoon{{{\cop \cdot
    p_k}}\log p_k}$
    & $\Cpl{\cop}{}$
    &     $p$& $-1$ & $1$ & $\cop$ & $-$\\
    &&&&&&\\
    \centering     $\left(\sum_{k=1}^{n} p_k^{1 + \cop \lambda}\right)^{1/\lambda}$
    &$\Cpl{\cop,\lambda}{}$
    & $p$& $-1$ & $1$ & $\cop$ & $\lambda$\\
    &&&&&&\\
    \centering Bhatia \cite{Bhatia_10}&
    &&&&&\\    &&&&&&\\
    \centering $\sum_{k=1}^n \twoon{{\dfrac{\cop \cdot p_k^\beta  \log p_k}{\sum_{k=1}^{n} p_k^\beta}}}$ &
    $\Cpl{\cop}{}$
    &     $p^\beta$& $-1$ & $1$ & $\cop$ & $-$\\
    &&&&&&\\
    \centering $\left(\dfrac{\sum_{k=1}^{n} p_k^{\beta + \cop \lambda}}{\sum_{k=1}^{n}
    p_k^\beta}\right)^{1/\lambda}$&
    $\Cpl{\cop,\lambda}{}$
    & $p^\beta$& $-1$ & $1$ & $\cop$ & $\lambda$\\
    \end{longtable}

\end{center}


\bibliographystyle{plain}
\bibliography{reference}

\begin{thebibliography}{10}

\bibitem{Abe_97}
Sumiyoshi Abe.
\newblock A note on the q-deformation-theoretic aspect of the generalized
  entropies in nonextensive physics.
\newblock {\em Physics Letters A}, 224(6):326 -- 330, 1997.

\bibitem{aczel1966lectures}
Joseph Acz{\'e}l.
\newblock {\em Lectures on functional equations and their applications}.
\newblock New York: Academic Press, 1966, edited by Oser, Hansjorg, 1966.

\bibitem{Aczel_Daroczy_63}
J.~Aczél and Z.~Daróczy.
\newblock {{"U}ber verallgemeinerte quasilineare {M}ittelwerte, die mit
  {G}ewichtsfunktionen gebildet sind}.
\newblock {\em Publ. Math. Debrecen}, 10:171–190, 1963.

\bibitem{Amblard_Vignat_06}
Pierre-Olivier Amblard and Christophe Vignat.
\newblock A note on bounded entropies.
\newblock {\em Physica A: Statistical Mechanics and its Applications},
  365(1):50--56, 2006.

\bibitem{Arimoto_71}
S.~Arimoto.
\newblock Information-theoretic considerations on estimation problems.
\newblock {\em Information and control}, 19:181--190, 1971.

\bibitem{Behara_Nath_74}
M.~Behara and P.~Nath.
\newblock Information and entropy of countable measurable partitions. i.
\newblock {\em Kybernetika}, 10:491--503, 1974.

\bibitem{Behara_Nath_80}
Minaketan Behara and Prem Nath.
\newblock On additive and non-additive measures of directed divergence.
\newblock {\em Kybernetika}, 16(1):(1)--12, 1980.

\bibitem{Bhatia_10}
P.K. Bhatia.
\newblock On certainty and generalized information measures.
\newblock {\em Int. J. Contemp. Math. Sciences}, 5(21):1035--1043, 2010.

\bibitem{Boekee_Lubbe_80}
D.E. Boekee and J.C.A.~Van der Lubbe.
\newblock The {R}-norm information measure.
\newblock {\em Information and Control}, 45(2):136 -- 155, 1980.

\bibitem{Borges_04}
Ernesto~P. Borges.
\newblock A possible deformed algebra and calculus inspired in nonextensive
  thermostatistics.
\newblock {\em Physica A: Statistical Mechanics and its Applications},
  340(1):95--101, 2004.

\bibitem{Daroczy_70}
Z.~Dar\'oczi.
\newblock Generalized information functions.
\newblock {\em Information and Control}, 16:36–51, 1970.

\bibitem{Ebanks_84}
B.~R. Ebanks.
\newblock Polynomially additive entropies.
\newblock {\em Journal of Applied Probability}, 21(1):179--185, March 1984.

\bibitem{Frank_Daffertshofer_00}
T.D. Frank and A.~Daffertshofer.
\newblock Exact time-dependent solutions of the renyi fokker–planck equation
  and the {F}okker–{P}lanck equations related to the entropies proposed by
  sharma and mittal.
\newblock {\em Physica A: Statistical Mechanics and its Applications},
  285(3):351--366, 2000.

\bibitem{Gupta_Sharma_76}
H.C. Gupta and B.D. Sharma.
\newblock On non-additive measures of inaccuracy.
\newblock {\em Czechoslovak Mathematical Journal}, 26:584--595, 1976.

\bibitem{Hardy_et_al_34}
G.~H. Hardy, G.~Polya, and J.~E. Littlewood.
\newblock {\em Inequalities, by G.H. Hardy, J.E. Littlewood [and] G. Polya}.
\newblock The University press, Cambridge [Eng.], 1934.

\bibitem{Havrda_Charvat_67}
Jan Havrda and Franti\v sek Charv\'at.
\newblock Quantification method of classification processes: {C}oncept of
  structural a-entropy.
\newblock {\em Kybernetika}, 1967.

\bibitem{Kaniadakis_02a}
G.~Kaniadakis.
\newblock {Statistical mechanics in the context of special relativity}.
\newblock {\em Physical Review E}, 66, 2002.

\bibitem{Kaniadakis_05}
G.~Kaniadakis, M.~Lissia, and AM~Scarfone.
\newblock Two-parameter deformations of logarithm, exponential, and entropy: a
  consistent framework for generalized statistical mechanics.
\newblock {\em Physical Review E}, 71(4):046128, 2005.

\bibitem{Kannappan_Rathie_74}
Palaniappan Kannappan and Pushpa~Narayan Rathie.
\newblock On a generalized directed-divergence function.
\newblock {\em Czechoslovak Mathematical Journal}, 24(1):5--14, 1974.

\bibitem{Kapur_83}
J.N. Kapur.
\newblock A comparative assessment of various measures of entropy.
\newblock {\em Journal of Information \& Optimization Sciences}, 4:207--232,
  1983.

\bibitem{Kerridge_61}
D.~F. Kerridge.
\newblock {Inaccuracy and Inference}.
\newblock {\em Journal of the Royal Statistical Society. Series B
  (Methodological)}, 23(1), 1961.

\bibitem{Khan_Autar_79}
Abul~Basar Khan and Ram Autar.
\newblock On useful information of order $\alpha$ and type $\beta$.
\newblock {\em Soochow Journal of Mathematics}, 5(December):93--99, 1979.

\bibitem{Mittal_75}
D.P. Mittal.
\newblock On some functional equations concerning entropy, directed divergence
  and inaccuracy.
\newblock {\em Metrika}, 22(1):35--45, 1975.

\bibitem{Nath_68}
P.~Nath.
\newblock On the measures of errors in information.
\newblock {\em Journal of Math. Sciences}, 3(1):1--16, 1968.

\bibitem{Nath_70}
P.~Nath.
\newblock An axiomatic characterization of inaccuracy for discrete generalized
  probability distributions.
\newblock {\em Opsearch}, 7:115--133, 1970.

\bibitem{Nath_68b}
Prem Nath.
\newblock Entropy, inaccuracy and information.
\newblock {\em Metrika}, 13(1):136--148, 1968.

\bibitem{Nivanen_Wang__03}
L.~Nivanen, A.~Le~M\'{e}haut\'{e}, and Q.~A. Wang.
\newblock {Generalized algebra within a nonextensive statistics}.
\newblock {\em Reports on Mathematical Physics}, 52(3):437--444, December 2003.

\bibitem{Onicescu_66}
O.~Onicescu.
\newblock Energie informationnelle.
\newblock {\em C. R. Acad. Sci. Paris Ser. A}, (263):841--842, 1966.

\bibitem{Pardo_Taneja_93}
L.~Pardo and I.J. Taneja.
\newblock Information energy and its aplications.
\newblock volume~80 of {\em Advances in Electronics and Electron Physics},
  pages 165 -- 241. Academic Press, 1991.

\bibitem{Pardo_86}
Leandro Pardo.
\newblock Order-$\alpha$ weighted information energy.
\newblock {\em Information Sciences}, 40(2):155 -- 164, 1986.

\bibitem{picard_79}
C.F. Picard.
\newblock Weighted probabilistic information measures.
\newblock {\em J. Inform. and Syst. Sci.}, 4:343--356, 1979.

\bibitem{Rathie_70}
P.N. Rathie.
\newblock On a generalized entropy and a coding theorem.
\newblock {\em J. Appl. Probl.}, 7:124--133, 1970.

\bibitem{Renyi_70}
Alfred Renyi.
\newblock {\em Probability Theory}.
\newblock Dover Publications, dover ed (2007) edition, May 1970.

\bibitem{shannon_48}
Claude~E. Shannon.
\newblock A mathematical theory of communication.
\newblock {\em The Bell System Technical Journal}, 27:379--423, 623--656, July,
  October 1948.

\bibitem{Sharma_Mittal_75}
B.D. Sharma and D.P. Mittal.
\newblock New non-additive measures of entropy for discrete probability
  distributions.
\newblock {\em Journal of mathematical sciences}, 10:28--40, 1975.

\bibitem{Sharma-Taneja_74}
BhuDev Sharma and InderJeet Taneja.
\newblock On axiomatic characterization of information-theoretic measures.
\newblock {\em Journal of Statistical Physics}, 10(4):337--346, 1974.

\bibitem{Singh_et_al_03}
R.P. Singh, Rajeev Kumar, and R.K. Tuteja.
\newblock Application of {H}older's inequality in information theory.
\newblock {\em Information Sciences}, 152(0):145 -- 154, 2003.

\bibitem{Taneja_05}
Inder~Jeet Taneja.
\newblock On symmetric and nonsymmetric divergence measures and their
  generalizations.
\newblock volume 138 of {\em Advances in Imaging and Electron Physics}, pages
  177 -- 250. Elsevier, 2005.

\bibitem{Taneja_Kumar_04}
Inder~Jeet Taneja and Pranesh Kumar.
\newblock Relative information of type s, csiszár's f-divergence, and
  information inequalities.
\newblock {\em Information Sciences}, 166(1--4):105 -- 125, 2004.

\bibitem{Theodorescu_77}
A.~Theodorescu.
\newblock Energie informationnelle et notions apparentees.
\newblock {\em {T}rabajos de {E}stadistica {Y} de {I}nvestigacion {O}perativa},
  28(2-3):183--206, 1977.

\bibitem{tsallis_88}
Constantino Tsallis.
\newblock Possible generalization of {B}oltzmann-{G}ibbs statistics.
\newblock {\em J. Statist. Phys.}, 52(1-2):479--487, 1988.

\bibitem{Tuteja_et_al_93}
R.~K. Tuteja, Shobha Chaudhary, and Priti Jain.
\newblock Weighted entropy of an order $\alpha$ and type $\beta$ information
  energy.
\newblock {\em Soochow Journal of Mathematics}, 19(2):129--138, 1993.

\bibitem{Lubbe_et_al_84}
J.~C.~A. Van Der~Lubbe and D.~E. Boekee.
\newblock A generalized class of certainty and information measures.
\newblock {\em Inf. Sci.}, 32(3):187--215, June 1984.

\bibitem{Varma_66}
R.S. Varma.
\newblock Generalizations of {R}\'enyi's entropy of order $\alpha$.
\newblock {\em {J}ournal of {M}athematical {S}ciences}, pages 34 -- 48, 1966.

\bibitem{Weaver_66}
Warren Weaver.
\newblock Probability, rarity, interest, and surprise.
\newblock {\em Pediatrics}, 38(4):667, 1966.

\end{thebibliography}

\end{document}